\documentclass[
 amsmath,amssymb,
 aps,
pra,
]{revtex4-2}

\usepackage{graphicx}
\usepackage{dcolumn}
\usepackage{bm}

\begin{document}

\preprint{APS/123-QED}

\title{Probing Spin Wave Diffraction Patterns of Curved Antennas}

\author{L. Temdie}%
 \altaffiliation[Also at ]{Lab-STICC - UMR 6285 CNRS, Technopole Brest-Iroise CS83818, 29238 Brest Cedex 03}
 \author{V. Castel}%
 \altaffiliation[Also at ]{Lab-STICC - UMR 6285 CNRS, Technopole Brest-Iroise CS83818, 29238 Brest Cedex 03}
\affiliation{ 
IMT Atlantique, Dpt. MO, Lab-STICC - UMR 6285 CNRS, Technopole Brest-Iroise CS83818, 29238 Brest Cedex 03}

\author{M.B. Jungfleisch}
\affiliation{University of Delaware, Dept. of Physics and Astronomy, Newark, Delaware, USA
}%
\author{R. Bernard}
\author{H. Majjad}
\author{D. Stoeffler}
\author{Y. Henry}
\author{M. Bailleul}
\affiliation{Université de Strasbourg, CNRS, Institut de Physique et Chimie des Matériaux de Strasbourg, UMR 7504, F-67000 Strasbourg, France}

\author{V. Vlaminck}
 \email{vincent.vlaminck@imt-atlantique.fr}
 \altaffiliation[Also at ]{Lab-STICC - UMR 6285 CNRS, Technopole Brest-Iroise CS83818, 29238 Brest Cedex 03}
\affiliation{ 
IMT Atlantique, Dpt. MO, Lab-STICC - UMR 6285 CNRS, Technopole Brest-Iroise CS83818, 29238 Brest Cedex 03} 


\date{\today}

\begin{abstract}
We report on the dependence of curvilinear shaped coplanar waveguides on the near-field diffraction patterns of spin waves propagating in perpendicularly magnetized thin films. Implementing the propagating spin waves spectroscopy techniques on either concentrically or eccentrically shaped antennas, we show how the link budget is directly affected by the spin wave interference, in good agreement with near-field diffraction simulations. This work demonstrates the feasibility to inductively probe a magnon interference pattern with a resolution down to 1$\mu$m$^2$, and provides a methodology for shaping spin wave beams from an antenna design. This methodology is successfully implemented in the case study of a spin wave Young's interference experiment.

\end{abstract}

\maketitle



\section{\label{sec:level1}Introduction}


The collective excitations of a spin ensemble, known as spin waves (or magnons for their quanta) \cite{Gurevich}, draw substantial interest as potential information carriers for unconventional electronic applications \cite{Chumak2017,Csaba2017,Barman2021,Pirro2021}. The versatility of the magnon dispersion in the broad microwave range offers a vast field of exploration for the development of wave-based computing technologies\cite{Chumak-book,Chumak2022,Mahmoud2022}, in which information could be encoded in both the phase and the amplitude of the spin wave. The manifold of nonlinear mechanisms along with the nanoscale integrability \cite{VVKruglyak2010} makes it a system of choice for the development of novel architectures such as neuromorphic computing \cite{Grollier2020,Papp2021-neural}, reservoir computing \cite{Papp2021,Allwood2023}, holographic memory \cite{Khitun2013,Khitun2015}, or spectral analysis \cite{Papp2017}, which are all interference-based techniques. Furthermore, the wide variety of non-reciprocal effects inherent to spin dynamics \cite{Gladii2016,Otlora2016,Grassi2020,Liu2018,Temdie2023, Temdie2023_MDPI} generates considerable interest for reducing the dimensions of analog signal processing components such as microwave isolators, circulator, filters, directional couplers, and phase shifters.  \\
Recently, basic concepts of optics applied to spin waves revealed the possibility of shaping and steering spin-wave beams in the sub-micron scale \cite{Stigloher2016,Gruszecki2017,Loayza2018,Grfe2020}, opening up new perspectives for the development of interferometric magnonic devices. Along these efforts, we developed a robust model to map the near-field diffraction pattern of arbitrary shaped antennas \cite{Vlaminck2023}, which allows to comprehend the magnon beamforming in extended thin films as a result of the excitation geometry.\\
In this article, we experimentally probe via out-of-plane spin wave spectroscopy the diffraction pattern of curvilinear antenna. The manuscript is organized as follow: In section~\ref{sec:level2}, 
we present a comparative study of spin wave transduction between straight and concentric pairs of coplanar waveguides. In section ~\ref{sec:level3}, we study a geometry of antenna that is akin to a Young's interference experiment for spin-waves. The design of these experiments relies on the near-field diffraction (NFD) simulation \cite{Vlaminck2023}, which was proven to benchmark spin wave diffraction in thin films for arbitrary excitation geometries.



\begin{figure*}[t]
\centering
\includegraphics[width=17cm]{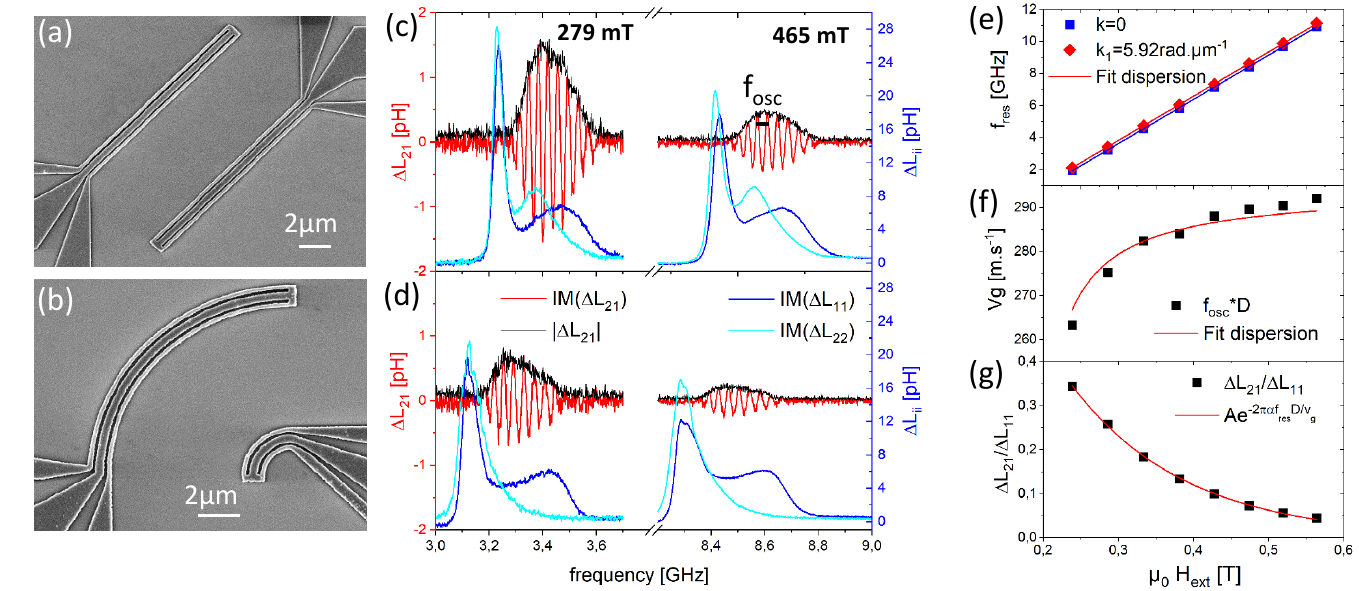}
\caption{(\textbf{a}) SEM image of a pair of identical straight antennas. (\textbf{b}) SEM image of a concentric antennas device. Propagating spin wave spectra measured at 279mT and 465mT respectively for the straight antennas (\textbf{c}), and concentric antenna (\textbf{d}). Magnetic characterization done on the straight antenna device. Field dependence of: (\textbf{e}) the resonance frequency of both $k\approx0$ and $k_1$ modes, (\textbf{f}) the measured group velocity, (\textbf{g}) the spectra amplitude.  \label{fig1}}
\end{figure*}

\section{\label{sec:level2}Concentric vs straight antennas}
\subsection{Sample fabrication and measurement protocol}
We firstly compare the transduction of spin-waves between pairs of identical straight antennas with the one of quarter circular concentric antennas, for which we kept the same separation distance $D$, and the same length of excitation antenna, namely $L_{ant}=\frac{\pi}{2}\,R\approx15.7\mu$m ($R$=10\,$\mu$m). Fig.~\ref{fig2}-(a),(b) show SEM images of two such antennas devices with a separation distance of $D$=8\,$\mu m$, which consist in Au-coplanar waveguide (CPW) with the following dimensions: a central line of $S$=400\,nm width, a ground lines of $G$=200\,nm width, spaced by 200\,nm. These dimensions of CPW produce wave packet centered around $k_1\approx$6\,rad.$\mu$ m$^{-1}$ \cite{Vlaminck2023}. The antennas were fabricated on top of an extended 30\,nm-thin sputtered Yttrium Iron Garnet (YIG) film \cite{Jungfleisch2016} via ebeam lithography, followed by lift-off of 5\,nm Ti/60nm Au. A 40\,nm $SiO_2$ spacer was deposited on top of the YIG film prior to the process. For thise study, we also fabricated similar comparative devices with a separation distance $D$=5\,$\mu$m.\\
The sample is placed directly onto the pole of a vertical electromagnet that can reach up to 1.3\,T at 5\,A, and contacted via 150\,$\mu$m-pitch picoprobe to an Agilent E8342B-50GHz vector network analyzer. We proceed to spin wave spectroscopy measurement at constant applied field sweeping the frequency in the [1-12]\,GHz range. In order to resolve a zero base line, we always subtract reference spectra acquired at different applied values ($H_{ref}$), for which no resonant feature occurs within the frequency sweep. Besides, we convert the $S_{ij}$ matrix to the impedance matrix $Z_{ij}$, which we divide by $i\omega$ to represent our spectra in units of inductance, accordingly with the inductive nature of the coupling between a spin wave and a coplanar waveguide \cite{Bailleul2003,Vlaminck2010}:
\begin{equation}
\Delta\,L_{ab}(f,H)=\frac{1}{i\omega}(Z_{ab}(f,H)-Z_{ab}(f,H_{ref}))
\label{DeltaL}
\end{equation}
where the subscripts ($a,b$) denote either a transmission measurement from ports b to port a, or a reflection measurement done on the same port if a=b.\\


\subsection{Spin wave spectroscopy}
Fig.~\ref{fig1}-(c),(d) shows reflection (blue) and transmission (red) spectra obtained at 279\,mT and 465\,mT and an input power of -15\,dBm, respectively, for a pair of straight antennas (upper panel), and for a pair of concentric antennas (lower panel), both with a separation distance of 8\,$\mu$m. We identify from the reflection spectra two main peaks. The first peak has a larger amplitude and appears at lower frequency. It corresponds to the FMR peak ($k\approx0$), namely, the region of the CPW extending from the 150 $\mu$m-pitch picoprobe contacts to the slightly reduced section of the CPW, yet wider than 10\,$\mu$m. The second peak corresponds to the $k_1$ sub-micron termination of the CPW shown in Fig.~\ref{fig1}-(a),(b), where microwave power is transmitted from port 1 to port 2 via spin-waves. One notices in particular the seeming lack of reflection peak $\Delta L_{22}$ for the 2\,$\mu$m-radius circular probe antenna (lower panel Fig.~\ref{fig1}-(d)), accordingly with the proportionality of the signal amplitude with the length of the antenna. \\
The transmission spectra reveal the typical features of propagating spin wave spectra \cite{Vlaminck2010,Loayza2018}, namely oscillations convoluted with an envelope, i.e. black and red lines in Fig. 2(b,c), respectively. One notices that the envelope appears less symmetrical with respect to frequency for the concentric geometry than for the straight one, which is likely due to interferences caused by the near-field diffraction pattern of the concentric antenna.

\begin{figure*}[t]
\centering
\includegraphics[width=18cm]{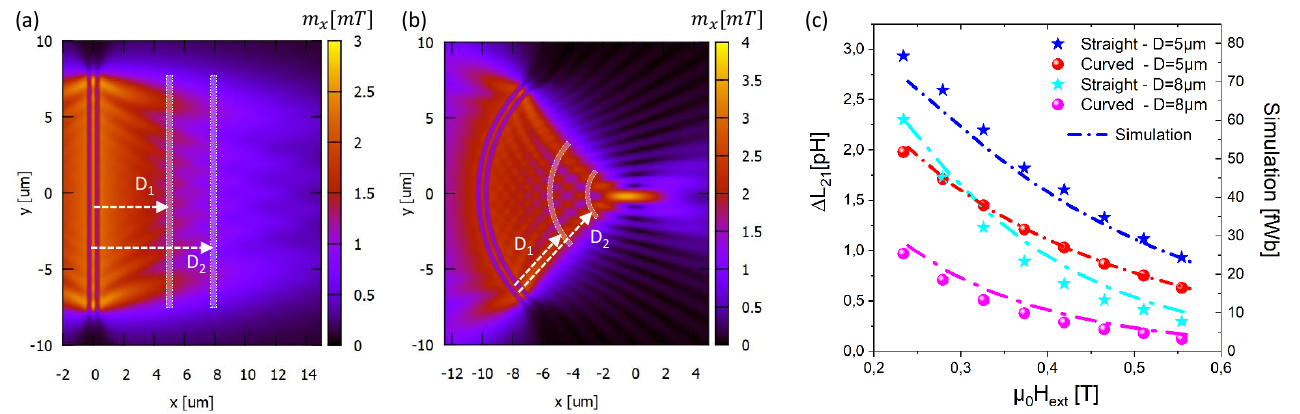}
\caption{NFD simulations at $\mu_0H_{ext}$=419.2\,mT, $f_{exc}$=6.892\,GHz for (\textbf{a}) a $15.7\mu$m-long straight antenna, and (\textbf{b}) 10$\mu$m-radius quarter circular antenna. (\textbf{c}) Comparison of the spin wave amplitude straight vs concentric with the emulated inductive signal from the NFD simulations. \label{fig2}}
\end{figure*}  

More importantly, we observe a clear diminution of amplitude for the concentric geometry compared with the straight one, with a rather constant ratio $\frac{\Delta L_{21}^{concentric}}{\Delta L_{21}^{straight}}\approx0.43$ over the whole frequency range. This observation may seem surprising at first, considering the confined nature of the radiation pattern with respect to the probe antenna definition (cf Fig.~\ref{fig2}-(b)), and knowing that we kept the same length of antenna for the excitation, and the same separation distance for both geometries. 
Spin wave dispersion in out-of-plane magnetized films are known to be isotropic, and considering that equal amount of power radiates inwards or outwards from the circular antenna, one might expect a comparable amplitude between straight and concentric geometries. \\
However, one can grasp this difference of amplitude by making an analogy with the Friis transmission formula used in telecommunications engineering \cite{Friis1946}, which relates received and emitted powers between two radio antennas to the product of their effective aperture area, accordingly with the concept of directivity for an antenna having uniform and equiphase aperture. In our case of spin-wave propagating in 2D, this analogy would give an amplitude ratio proportional to the square-root of the ratio of the arc lengths: $\frac{\Delta L_{21}^{concentric}}{\Delta L_{21}^{straight}}\propto\sqrt{\frac{R_2}{R_1}}\approx0.45$. Still, this agreement should be viewed cautiously as the Friis formula is normally applicable in the far-field region to ensure a plane wave front at the receiving antenna, which corresponds here to a propagation distance $D\geq \frac{(\pi R_1)^2}{\lambda}\approx$1\,mm. For this reason, we ought to resort to near-field diffraction simulations in order to assess the conformity of our measurements.    \\
We now present in Fig.~\ref{fig1}-(e)-(g) the methodology used to evaluate the magnetic properties used in the near-field diffraction simulations, from the spin wave spectroscopy performed over the whole field range on a single pair of straight antenna, for which we can ensure a plane wave profile.   
Firstly, we track the field dependence of the $k=0$ reflection peak and the transmission peak as shown in Fig.~\ref{fig1}-(e), and fit it to the MSFVW dispersion relation \cite{Kalinikos1981}, which gives a gyromagnetic ratio $\frac{\gamma}{2\pi}$=28.2$\pm$0.1\,GHz.T$^{-1}$, an effective magnetization $\mu_0M_s$=185$\pm$5\,mT more or less equal to the saturation magnetization, suggesting no uniaxial anisotropy for our YIG film. We then estimate the group velocities $v_g$ from the period of oscillation of the transmission spectra \cite{Vlaminck2010,Loayza2018}, and fit its field dependence to the dispersion relation as shown in Fig.~\ref{fig1}-(f), letting only the exchange constant as a free parameter, which gives $A_{exch}$=3.5$\pm$0.2\,pJ.m$^{-1}$.
Finally, we fit the field dependence of the transmission amplitude to an exponential decay $\Delta L_{21}\propto exp(-D/L_{att})$ Fig.~\ref{fig1}-(g), for which we adopted the low wavevector approximation of the attenuation length $L_{att}=\frac{v_g}{2\pi \alpha f_{res}}$, where $f_{res}$ is expressed from the Kalinikos-Slavin expression \cite{Kalinikos1981}. We obtain a Gilbert damping of $\alpha$=9.1$\pm$0.5 10$^{-4}$, which appears slightly bigger than previously reported values on similar sputtered thin YIG films \cite{Loayza2018,Li2016}. We note that the same methodology applied to the $D$=5\,$\mu$m straight antennas device gives very close results within the estimated error bars. 

\begin{figure*}[t]
\centering
\includegraphics[width=18cm]{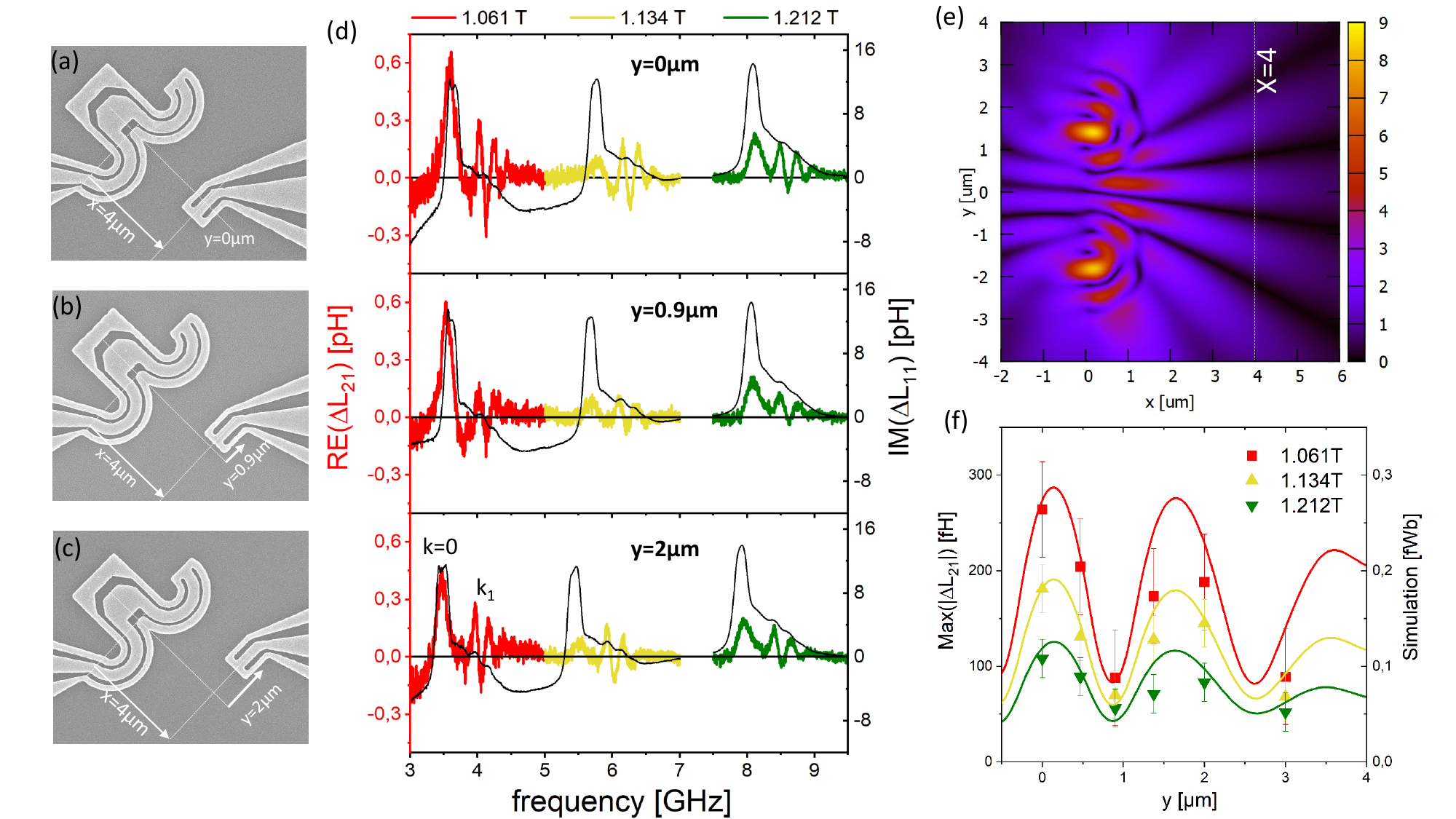}
\caption{(\textbf{a})-(\textbf{c}) SEM images of three Young's interference devices with different location of probe antenna. (\textbf{d}) Transmission (left y-axis), and reflection spectra (right y-axis) for the corresponding devices.  (\textbf{e}) NFD simulation of a Young's interference antenna performed at $\mu_0$H$_{ext}$=1.061\,T and f=4.07\,GHz. (\textbf{f}) Comparison of the evolution of the spin wave amplitude at $x$=4\,$\mu$m with the emulated inductive signal from the NFD simulations (color palette in unit of mT).\label{fig3}}
\end{figure*}  

\subsection{Comparison with NFD simulations}

In order to assess the conformity of our measurements, we performed near-field diffraction (NFD) simulations \cite{Vlaminck2023} for each field values respectively for the straight and the circular excitation antenna.    
Fig.~\ref{fig2}-(b),(c) show the simulated magnetization amplitude expressed in units of mT for an YIG film magnetized out-of-the plane with an external field $\mu_0H_{ext}$=419.2\,mT, and for an excitation frequency of $f_{exc}$=6.892\,GHz, respectively for a $\frac{\pi}{2}\,R$-long straight CPW, and a quarter-circular CPW with radius of curvature $R$=10\,$\mu m$. Both CPWs have the same lateral dimensions, namely a central line $w_s$=400\,nm and a ground line $w_g$=200\,nm, correspondingly with the measured devices. We defined the microwave magnetic field components $(h_x,h_y)$ from the Oersted field of a straight conductor with rectangular section carrying uniform current density, whose value was adjusted according to typically used input power and impedance of the antennas. We note that the field distribution obtained with this somewhat crude approximation compares very well with finite element simulations of curved coplanar waveguides \cite{Vlaminck2023}.\\

 In order to compare the simulations with propagating spin wave spectra obtained from several pairs of antennas, we perform a sum over an effective area where the detection antenna is located as represented in white on the simulations of Fig.~\ref{fig2}-(b),(c), and multiplied by the pixel area $dxdy$. Indeed, the coupling of a spin-wave with a CPW, which is inductive in nature, can be estimated by the magnetic flux sensed by the antenna.
 Although it does not strictly correspond to the dynamic field radiated from the spin wave that the probe antenna senses, this rather simple averaging method provides a comprehensive estimate of the antenna's shape-dependent transduction, which we express in units of magnetic flux, e.g. in femto Weber (fWb).
 Fig.~\ref{fig2}-(c) summarizes the field dependence of all measured transmission amplitudes ($\Delta L_{21}$-left y-axis) compared with the emulated inductive signal from the NFD simulations (right y-axis) for both the straight and the concentric pairs of antennas. We find an excellent matching between the amplitude of the measured transmission spectra and the simulated inductive signal over the whole field range for both the $D$=5\,$\mu$m and the $D$=8\,$\mu$m series. The agreement between measurements and simulations is all the better here that the antenna design matches with the confined diffraction pattern, e.g. no spin-wave dynamics is to be found in the transition to the probe antenna's termination. This explanation of the differences in link budget between concentric and straight pairs of antennas validates our understanding of spin wave transduction in curvilinear geometries in terms of near-field diffraction.\\
Furthermore, the ratios of amplitude $(\frac{\Delta L_{21}^{concentric}}{\Delta L_{21}^{straight}})$ remains fairly constant and close to the square root of the radius ratio $\sqrt{\frac{R_2}{R_1}}$, namely $(\frac{\Delta L_{21}^{concentric}}{\Delta L_{21}^{straight}})_{5\,\mu m}\approx 0.66\pm0.01$ and $(\frac{\Delta L_{21}^{concentric}}{\Delta L_{21}^{straight}})_{8\,\mu m}\approx 0.43\pm0.01$. The comparison is better for the longer separation distance as suggested by the analogy with the Friis formula, which only applies in the far-field region.

\section{\label{sec:level3}Young's interference experiment}
We explore here the idea of magnon beamforming from the shape of an excitation antenna, and propose to reiterate a Young's interference experiment with two seemingly circular apertures. Fig.~\ref{fig3}-(a)-(c) show scanning electron (SEM) images of the Young's interference devices consisting in two adjacent semi-circular 1\,$\mu$m wavelength CPW, having each a 1\,$\mu$m curvature radius for the central line, and whose centers are 2\,$\mu$m apart. We fabricated a series of 6 such devices on top of a 20\,nm-thin Ni$_{80}$Fe$_{20}$ film, changing the location of the probe antenna, namely, keeping the same $x=4\,\mu$m and varying $y$=[0.0,0.47,0.9,1.375,2.0,3.0]. In this manner, we can perform a discrete mapping of this Young's interference pattern with sub-micron resolution, using a 1\,$\mu$m$^2$ square CPW as probe antenna. The tightness of the aimed curvature could not allow to fabricate this sub-micron size geometry on a YIG film, due to the limitations posed by the conductive resine \cite{Electra}. \\
 Fig.~\ref{fig3}-(d) shows the transmission spectra $\Delta L_{21}$ (colored lines, left y-axis) and reflection spectra $\Delta L_{11}$ (black line, right y-axis) for the three devices with a probe antenna position at $y=0$ for the device shown in Fig.~\ref{fig3}-(a), $y=0.9\,\mu$m for the one of Fig.~\ref{fig3}-(b), and $y=2\,\mu$m for the one of Fig.~\ref{fig3}-(c). All devices were measured at $-15$\,dBm input power, and for 3 different applied fields: 1.061\,T,1.134\,T, and 1.212\,T. The transmission spectra display a first peak at lower frequency, which should not be mistaken with a propagating spin wave signal, as it is aligned with the k=0 peak of the reflection spectra. Therefore, we focus on the remaining part of the spectra featuring the typical oscillations of the $k_1$ spin-wave mode, in order to track the change of amplitude with the probe antenna position. For the three field values, the oscillation amplitude appears maximum for the $y=0$ device, it is significantly reduced for the $y=0.9\,\mu$m device, while it increases again for the $y=2\,\mu$m device.\\
 We show in Fig.~\ref{fig3}-(e) a NFD simulation of this Young's interference device done at $\mu_0$H$_{ext}$=1.061\,T and f=4.07\,GHz, using the following set of parameters accordingly with prior characterization of this permalloy film \cite{PhDthesis_Vlam}: a saturation magnetization of $\mu_0M_{s}$=0.95\,T, a gyromagnetic ratio $\gamma$=29.8\,GHz.T$^{-1}$, and a Gilbert damping constant $\alpha$=7.5\,10$^{-3}$, and exchange constant $A_{exch}$=7.5\,pJ.m$^{-1}$. The diffraction pattern shows clearly the formation of spin wave beams separated by dark zones, corresponding respectively to the constructive and destructive interference of spin waves in a similar fashion as a double-slit experiment in optics. \\
We finally compare in Fig.~\ref{fig3}-(f) the transmission spectra amplitude obtained on the 6 devices with the emulated inductive signal from the corresponding NFD simulations as described in sec.\ref{sec:level2}. We obtain a satisfying agreement, reproducing on one hand the spatial dependence of the spin wave diffraction pattern over two constructive interference beams, and on the other hand the comparative amplitude between the 3 field values. The little discrepancy between simulations and measurements could be due to the part of the CPW that transitions to the 1\,$\mu$m$^2$ termination, which can slightly pick up some flux within the remaining diffraction pattern. In essence, this study demonstrates the possibility to shape spin wave beams from the shape of an antenna, and resolve sub-micron featured-size diffraction pattern with a 1\,$\mu$m$^2$ inductive probe.

\section{\label{conclusion}conclusion}
We presented a study on the spin wave transduction from curved excitation antennas, comparing transmission spectras with simulated mappings of the spin wave amplitude for various geometries of excitation. We firstly showed that the difference in transmission amplitude between pairs of straight antenna versus concentric antennas was very well reproduced over a broad frequency range by an emulated inductive signal built from the NFD mapping combined with the probe antenna definition. This validates our understanding of spin wave transduction in curvilinear geometries in terms of near-field diffraction. Secondly, we reiterated a Young double-slit experiment with an antenna made of two adjacent semi-circular CPW, acting like two seemingly circular apertures. We satisfyingly reproduced the simulated spin wave diffraction pattern with a series of devices varying the position of probe antenna. We demonstrated in particular the possibility to inductively sense the spin wave amplitude with a 1\,$\mu$m$^2$ spatial resolution. 
These results provide a methodology to explore the magnon beamforming through the shape of an excitation antenna, and pave the way for future development of interferometric magnonic sensors.

\begin{acknowledgments}
The authors would also like to acknowledge the financial support from the French National research agency (ANR) under the project \textit{MagFunc}, the Département du Finistère through the project \textit{SOSMAG}, and also the Transatlantic Research Partnership, a program of FACE Foundation and the French Embassy under the project \textit{Magnon Interferometry}.
\end{acknowledgments}

\nocite{*}

\bibliography{SWTCA_V2}

\end{document}